\documentclass[aps,reprint,prl]{revtex4-1}
\usepackage[dvipsnames]{xcolor}

\usepackage{bm}
\usepackage{graphicx} %
\usepackage{amsmath} %
\usepackage{amssymb} %
\usepackage{url}
\usepackage{subfigure} %
\usepackage{float} %
\usepackage{upgreek} %
\usepackage{multirow} %
\usepackage{color} %
\usepackage{lineno} %
\usepackage{hyperref} %
\usepackage{booktabs}
\usepackage{placeins}

\hypersetup{colorlinks=true, urlcolor=black, linkcolor=blue, citecolor=red} %

\usepackage{ulem}
\allowdisplaybreaks[4]

\begin{document}

\title{Cosmology with the thermal-kinetic Sunyaev-Zel'dovich effect}

\author{William Coulton${}^1$}
\email{wcoulton@ast.cam.ac.uk}
\author{ Atsuhisa Ota${}^2$}
\email{a.ota@damtp.cam.ac.uk}
\author{Alexander van Engelen${}^{3,4}$}
\email{alexander.van.engelen@asu.edu}
\affiliation {${}^1$Institute of Astronomy and Kavli Institute for Cosmology Cambridge, Madingley Road, Cambridge, CB3 0HA, UK}
\affiliation {${}^2$Department of Applied Mathematics and Theoretical Physics, University of Cambridge, Cambridge, CB3 0WA, UK }
\affiliation {${}^3$School of Earth and Space Exploration, Arizona State University, Tempe, AZ, 85287, USA} 
\affiliation{ ${}^4$Canadian Institute for Theoretical Astrophysics, University of Toronto, 60 St George St, Toronto, ON, M5S 3H8, Canada}

\date{\today}

\begin{abstract}
Compton scattering of the cosmic microwave background (CMB) from hot ionized gas produces a range of effects, and the leading order effects are the kinetic and thermal Sunyaev Zel'dovich (kSZ and tSZ) effects. In the near future, CMB surveys will provide the precision to probe beyond the leading order effects. In this work we study the cosmological information content of the next order term which combines the tSZ and kSZ effects, hereafter called the thermal-kinetic Sunyaev Zel'dovich (tkSZ) effect. As the tkSZ effect has the same velocity dependence as the kSZ effect, it will also have many of the useful properties of the kSZ effect. However, it also has its own, unique spectral dependence, which allows it to be isolated from all other CMB signals. We show that with currently-envisioned CMB missions the tkSZ effect can be detected and can be used to reconstruct large scale velocity fields, with no appreciable bias from either the kSZ effect or other extragalactic foregrounds. Furthermore, since the relativistic corrections arise from the well-studied pressure of ionized gas, rather than the gas number density as in the kSZ effect, the degeneracy due to uncertain gas physics will be significantly reduced. Finally, for a very low-noise experiment the tkSZ effect will be measurable at higher precision than the kSZ.


\end{abstract}

\maketitle


\textit{Introduction---.}
Cosmological observations will significantly increase in size and scope in the next few decades.
In particular, galaxy surveys and high resolution measurements of secondary anisotropies in the cosmic microwave background~(CMB) will shed light on new aspects of cosmology
~\cite{Ade:2018sbj,Abazajian:2016yjj,Abell:2009aa,Aghamousa:2016zmz}.
These efforts will be powerful themselves, but taking cross-correlations of these observables will provide us further information of the Universe.
One such strategy would be large scale velocity reconstruction with the kinetic Sunyaev-Zel'dovich~(kSZ) effect \cite{Sunyaev:1980nv}, which arises from bulk flows of free electrons along the line of sight. The first detections of the kSZ effect have been obtained in the past few years, using galaxy redshift surveys together with high-resolution maps of the CMB \cite{Hand:2012ui,DeBernardis:2016pdv,Soergel:2016mce,Hill:2016dta,Aghanim:2017bzn,Schaan2016, Ma:2017ybt}. The precision of these measurements will increase drastically in the near future \cite{Munchmeyer:2018eey,Ferraro:2016ymw,Deutsch:2017ybc,Pan:2019dax}. 
These measurements are probes of both large-scale bulk flows and the smaller-scale distribution of electrons in halos. Given an understanding of the latter, we can constrain the former and reconstruct large-scale velocities in the Universe. While this has significant possibilities for constraining fundamental cosmology, including modifications to general relativity, it is subject to astrophysical uncertainty arising from the distribution of ionized gas within dark matter halos. This uncertainty is known as the ``optical depth degeneracy''~(e.g. Ref.~\cite{Alonso:2016jpy}) and several methods have been put forward to reduce it \cite{Flender:2016cjy,Battaglia:2016xbi,Madhavacheril:2019buy}. In this article we explore the potential of relativistic corrections to the SZ effects~\cite{Challinor:1997fy,Challinor:1998sg,Itoh:1997ks,Nozawa:1998zu}, which we call the thermal kinetic Sunyaev-Zel'dovich~(tkSZ) effect, for velocity reconstruction~\footnote{We use the acronym of ``tkSZ'' to avoid confusion with rotational kinetic Sunyaev-Zel'dovich~(rkSZ) discussed in Refs.~\cite{Baxter:2019tze,Zorrilla2019}.}.
The tkSZ effect arises from the thermal pressure of electron gas in moving halos, whereas the kSZ arises from the gas density in moving halos; depending on the gas temperature it is about a 1\% correction to the kSZ effect. Its frequency dependence is different from both of the kSZ and thermal Sunyaev-Zel'dovich~(tSZ) effects. Given observations with enough frequency bands, we can distinguish and isolate each of these signals.
Using the recently proposed kSZ tomography technique~\citep{Smith:2018bpn}, we show for the first time that the tkSZ effect is detectable via the galaxy-galaxy-tkSZ bispectrum in currently envisioned observational projects such as the Probe of Inflation and Cosmic Origins~(PICO)~\cite{Hanany2019} or CMB-HD experiments~\cite{Sehgal2019} . This work complements Ref.~\cite{Remazeilles2019} which discussed the power spectrum of higher order thermal corrections to the tSZ effect.

\textit{Relativistic SZ effects---.}
The SZ effects are spectral distortions to the homogeneous and isotropic blackbody spectrum of the CMB.
They are produced when the CMB photons are scattered by electrons in halos after recombination, and hence they are useful to investigate the thermodynamics of the electrons, and the peculiar motions of the halos.
To compute the SZ effects, in principle, we have to solve a Boltzmann equation for photons including the Compton scattering collision effects.
The equation is a complicated partial differential and integral equation, which we can solve numerically.
However, for cosmology, we are interested in using the properties of many SZ sources, each of which can be thought of as a point source in the sky.
Hence, we need a more efficient way to compute the SZ effects for many halos.
The Boltzmann equation is simplified by the single scattering approximation and when the relativistic corrections are small.
With these assumptions, we apply the moment expansion schemes that were developed in Refs.~\cite{Challinor:1997fy,Itoh:1997ks,Nozawa:1998zu,Challinor:1998sg}.

Let $\nu$ and $T_{\rm CMB}=2.725$K be the photons' frequency and the CMB temperature today.
Then, we know the background phase space distribution function of the CMB is written by a Planck distribution $\bar f \equiv [\exp(h \nu/k_{\rm B}T_{\rm CMB})-1]^{-1}$, with $h$ and $k_{\rm B}$ being the Planck and Boltzmann constants.
We Taylor expand the CMB distribution function, $f$, around $\bar f$ in terms of the electron temperature-mass ratio $\theta_{\rm e} \equiv k_{\rm B}T_{\rm e}/ m_{\rm e} c^2$, and the halos' 3-velocity ${\bm v}$.
Defining the derivative operator $\mathcal D \equiv -\nu\partial/\partial \nu$, we introduce two functions of photon frequency $\nu$: $\mathcal G \equiv \mathcal D \bar f$, and $\mathcal Y  \equiv (\mathcal D^2 -3 \mathcal D)\bar f$.
$\mathcal G$ is the frequency dependence of the primary temperature anisotropies, and $\mathcal Y$ is that of the Compton $y$ parameter, which does not change the number of photons.
We may generalize the $n$-th derivatives of the Planck distribution in terms of $\nu$ as $\mathcal Y^{(n)} \equiv \mathcal D^{n-2} \mathcal Y$ for $n\geq 2$.
Using these functions, we may write the spectral distortion $\delta f \equiv f -\bar f$ due to the Compton scattering in halos as~\cite{Challinor:1997fy,Itoh:1997ks,Nozawa:1998zu,Challinor:1998sg}
\begin{align}
	\delta f =  \int \mathrm{\rm d}\chi n_{\rm e}\sigma_{\rm T} a e^{-\tau}S, \label{sol:SZ}
\end{align}
where $\tau$ is the optical depth, $\chi$, $n_{\rm e}$ and $a$ are the comoving distance, electron number density and scale factor in the CMB rest frame, and the source function is~\cite{Nozawa:1998zu,Challinor:1998sg,Chluba:2012dv}
\begin{align}
	&S= \theta_{\rm e}\mathcal{Y}^{(2)} 
	+\theta_{\rm e}^2 \left(-\frac{3}{10}  \mathcal{Y}^{(2)} -\frac{21}{10} \mathcal{Y}^{(3)} +\frac{7}{10} \mathcal{Y}^{(4)} \right)\notag \\
	&+
	 {\bm v}\cdot \mathbf n \mathcal{G}
	+  \theta_{\rm e} {\bm v}\cdot \mathbf n \left(\frac{2}{5} \mathcal{G}- \mathcal{Y}^{(2)} +\frac{7}{5} \mathcal{Y}^{(3)}  \right)
	\notag \\
	&+   \theta_{\rm e}^2{\bm v}\cdot \mathbf n\left(\frac{1}{5} \mathcal{G}-\frac{7}{10} \mathcal{Y}^{(3)} -\frac{33}{10} \mathcal{Y}^{(4)} +\frac{11}{10} \mathcal{Y}^{(5)} \right), \label{col:result:LI:2}
\end{align}
with line-of-sight direction $\mathbf n$. These terms correspond respectively to the tSZ effect, the relativistic thermal SZ effect (rtSZ), the kSZ effect, the tkSZ effect (which is the focus of this paper), and one higher order correction. We have dropped terms of $\mathcal O(v^2,\theta^3_{\rm e})$, which are known as subdominant contributions for typical halos~\cite{Nozawa:1998zu,Challinor:1998sg}.
We assumed that the incoming photon distributions to the halos are the homogeneous and isotropic Planck distribution.
Note that the optical depth in the CMB rest frame depends on the halos' velocity, and $-\mathcal Y^{(2)}$ in the second line is a frequency shift because of this effect~\cite{Chluba:2012dv}.
Eq.~\eqref{sol:SZ} evaluated on the sky can be factored into its dependence on line-of-sight direction and frequency according to
\begin{align}
	\delta f  (\nu,\mathbf n) = \Theta(\mathbf n)\mathcal G(\nu) +y(\mathbf n) \mathcal Y(\nu) +\alpha(\mathbf n) \mathcal A(\nu) +\cdots,\label{eff:Temp:def}
\end{align}
where $\Theta(\mathbf n)$ is a dimensionless temperature perturbation composed of the primary anisotropy, kSZ and integrated Sachs-Wolfe effects. $\mathcal A(\nu) \equiv 2\mathcal G (\nu)/5 - \mathcal{Y}^{(2)}(\nu) +7\mathcal{Y}^{(3)} (\nu)/5$ is the frequency response for the tkSZ effect parameterized by $\alpha(\mathbf n)$.
The dots imply the other effects.
Combining Eqs.~\eqref{sol:SZ} and \eqref{eff:Temp:def}, we obtain
\begin{align}
	\alpha =     \int \mathrm d\chi 
	   R p \frac{\bm v}{c} \cdot \mathbf n \label{def:alpha},\,\,\,
	   R \equiv \frac{a \sigma_{\rm T}}{m_{\rm e}c^2} e^{-\tau},
\end{align}
where $p\equiv n_{\rm e}k_{\rm B}T_{\rm e}$ is the electron pressure.
This effect can be contrasted with the standard tSZ effect $y = \int \mathrm d\chi R p$, and kSZ effect $\Theta_{\rm kSZ} = \int \mathrm d\chi a \sigma_T e^{-\tau} n_{\rm e} {\bm v}/c \cdot \mathbf n$.
Thus the calculation of the tkSZ effect is identical to that of the kSZ effect except the electron density is replaced by the pressure.
\begin{figure}[t]
  \centering
  \includegraphics[width=0.48\textwidth]{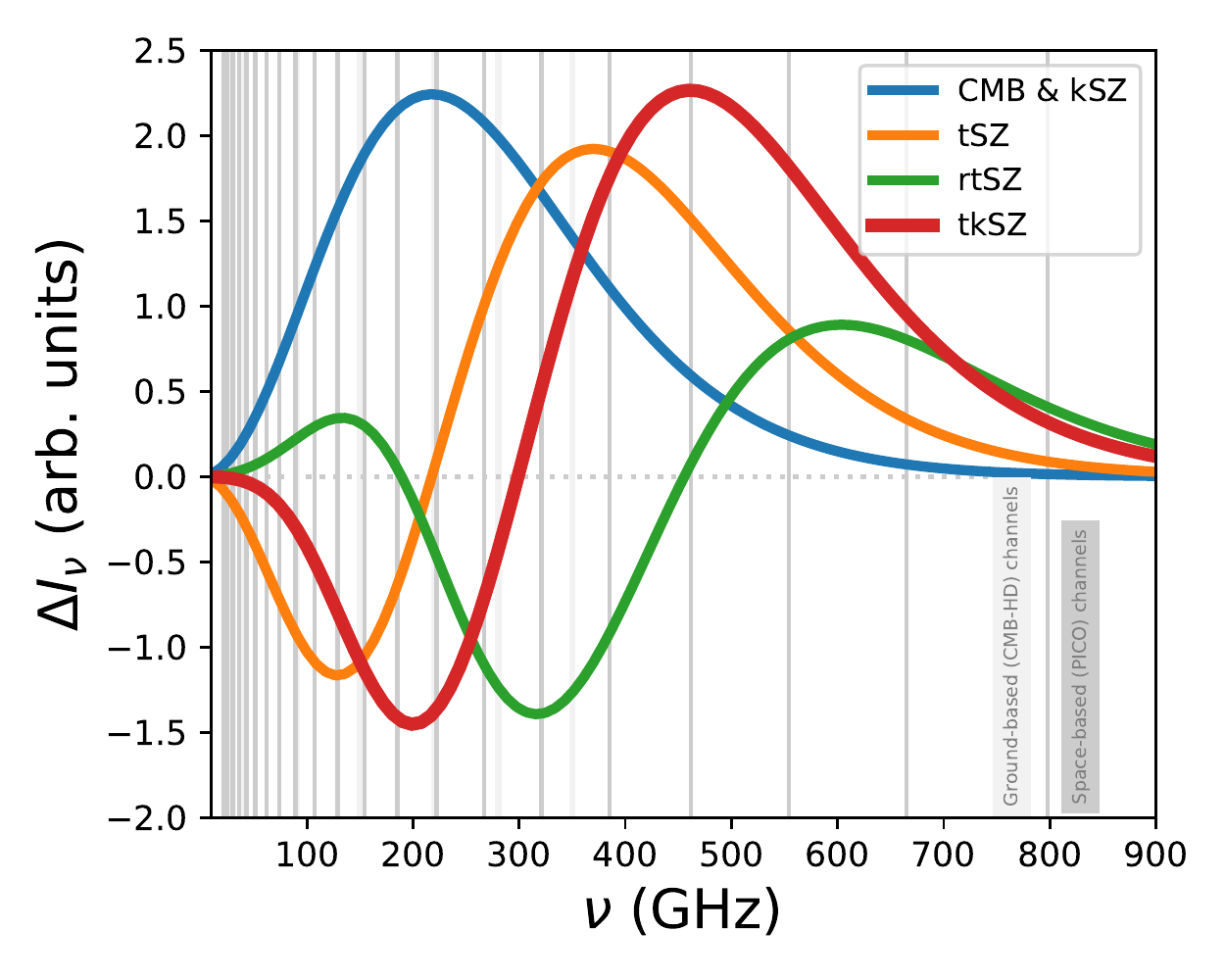}
\caption{Spectral response for several signals of interest. The kSZ effect, like the CMB temperature perturbation itself, is a shift of blackbody with temperature 2.7K (both blue, $\propto \nu^3\mathcal G  $). The tSZ effect is a spectral distortion that is negative for $\nu < 217$GHz and positive at $\nu > 217$GHz (orange, $\propto \nu^3\mathcal Y^{(2)}$). The tkSZ effect has a spectral dependence that peaks at higher frequencies (green, $\propto \nu^3\mathcal A$). Also shown is the term often referred to as the relativistic thermal SZ (red, $\propto \nu^3\mathcal B$) with $\mathcal B\equiv - 3\mathcal{Y}^{(2)}  /10  - 21\mathcal{Y}^{(3)}  /10 + 7\mathcal{Y}^{(4)}  /10 $. The frequency bands for the experiments we consider are shown in grey.} 
\label{fig:specResp}
\end{figure}

\textit{Isolating the signal---.}
The real sky is composed of many different signals, both galactic and extra-galactic. In order to study the much fainter tkSZ effect we need to isolate this signal from the other backgrounds.
Given observations with enough frequency channels we can use the unique spectral signature in Fig.~\ref{fig:specResp} to isolate the tkSZ effect from the other sky signals.\ We use the constrained internal linear combination (cILC) method~\cite{Remazeilles2011} to do this and to ensure that there is no contamination from the more significant kSZ effect with similar velocity dependence. Due to this velocity dependence, there is no bias to our detection methods (described below) from the tSZ or cosmic infrared background (CIB), but these signals will contribute effective noise. The cILC approach is designed to yield a minimum-noise map of the tkSZ effect with correct normalization, but with zero response to the kSZ; this is enabled by our perfect knowledge of the spectrum of both signals \footnote{We note that this method is blind to the spatial and frequency properties of the foregrounds, and parametric approaches, such as FGbuster~\cite{Stompor2016}, may offer greater detection prospects.}.  We briefly summarize this method here and refer the reader to Ref.~\cite{Remazeilles2011} for more details. 
The measured intensity at a frequency $\nu_i$, $s_{i}(\mathbf n)$, can be given by discretizing Eq.~\eqref{eff:Temp:def}, with the appropriate phase space volume element
\begin{align}
s_i(\mathbf n)  = \alpha(\mathbf n) 
A_i+\Theta(\mathbf n)  G_i+ N_i(\mathbf n),
\end{align}
where $ G_i\equiv 2h \nu_i^3 c^{-2}\mathcal G(\nu_i)$ is a vector of the spectral dependencies of the CMB anisotropies, and $ A_i\equiv 2h \nu_i^3 c^{-2}\mathcal A(\nu_i)$ is that of the tkSZ effect at the observed frequencies. The
$N_i(\mathbf n)$ are all other signals, which in this work we assume consist of the following components: the CMB, tSZ, late-time kSZ, reionization kSZ, rtSZ, CIB, radio galaxies, galactic dust emission, instrumental noise and, for ground-based surveys, atmospheric noise. In Table \ref{tab:foregrounds} we summarize the properties of these components, specifying their angular power spectra and spectral dependence. 
We assume that the CIB and tSZ are correlated at the 10$\%$ level \cite{Ade:2015ira} and the rtSZ, tSZ, tkSZ and kSZ correlations are described by the halo model. For simplicity we do not allow for frequency decoherence of the CIB. For the amplitude of the power spectrum of dust, we assume the measurement of Ref.~\cite{Dunkley2013}, which is a fit to a relatively small fraction of sky; however, we found that our results are unchanged if the amplitude of galactic dust power is increased by a factor of ten.

\begin{figure}[t]
  \centering
  \includegraphics[width=0.48\textwidth]{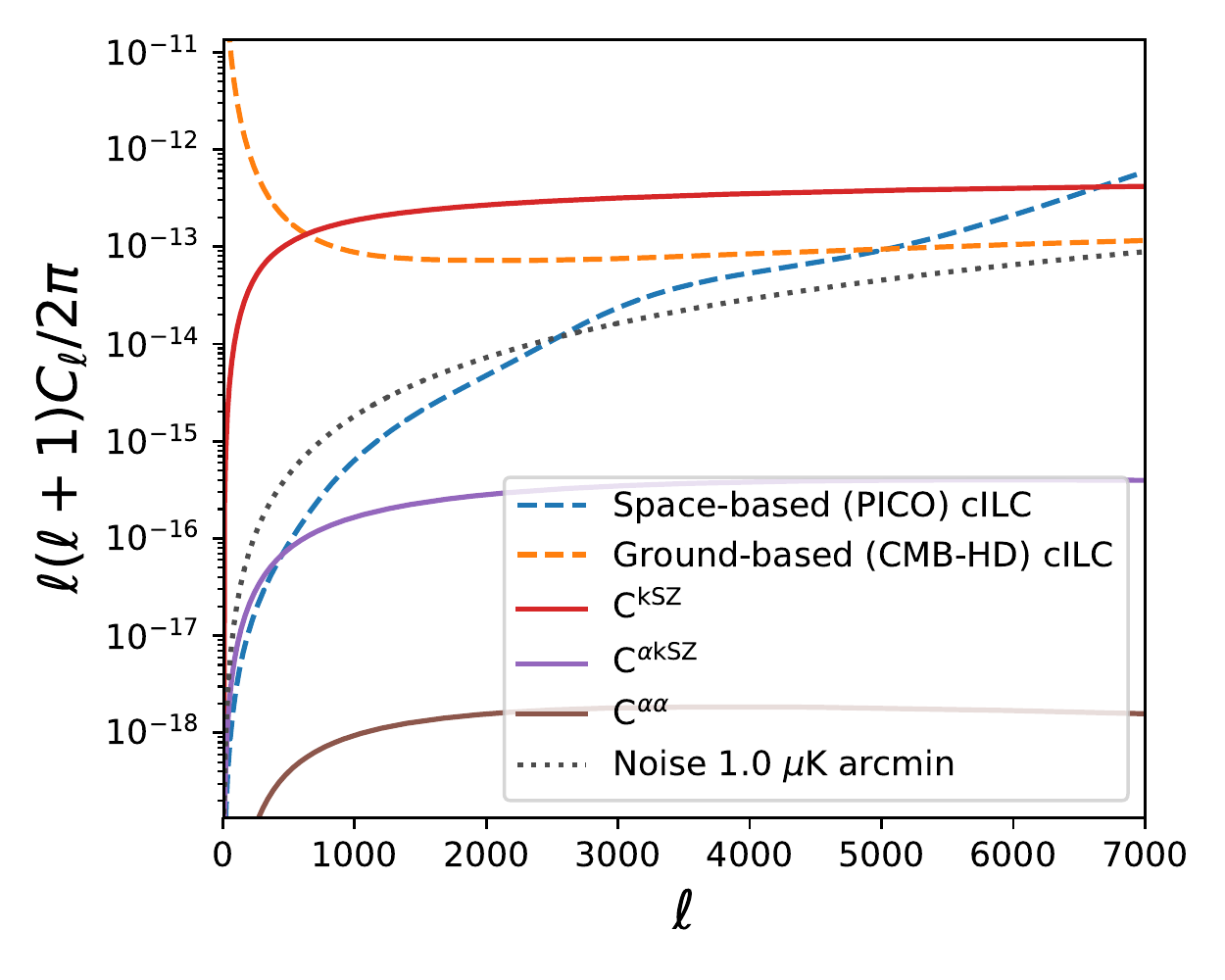} 
\caption{The tkSZ auto spectrum (brown) and its cross correlation with the kSZ effect (purple). Also plotted are the constrained ILC noise curves for the a CMB-HD (dashed orange) and PICO like experiment (dashed blue). For scale we also plot the kSZ power spectrum (red) along with the 1 $\mu$K-arcmin noise curve (dotted). }
\label{fig:powerspectra}
\end{figure}

We consider two reference experiments.  First, we consider a space-based survey with sufficient frequency resolution to separate a significant number of signals; our setup, shown in Table \ref{tab:pico}, is based on the recently proposed PICO experiment~\cite{Hanany2019}. Secondly, we consider a ground-based survey, with more modest frequency coverage but significantly higher angular resolution. For this survey we adopt the parameters of the recently-envisioned CMB-HD experiment~\cite{Sehgal2019}, detailed in Table ~\ref{tab:cmbhd}. 
We forecast the statistics of the isolated tkSZ map by using the following minimum variance estimator: 
\begin{align} \label{eq:cILCmap}
\hat{\alpha} =\frac{(\vec{ G}\cdot \Sigma^{-1}\vec{G}) \vec{ A} -(\vec{A}\cdot \Sigma^{-1}\vec{G})\vec{G} }{(\vec{A}\cdot  \Sigma^{-1} \vec{A}) ( \vec{G} \cdot \Sigma^{-1}\vec{G})  -(\vec{A} \cdot \Sigma^{-1} \vec{G}) ^2 } \cdot  \Sigma^{-1}\vec{s},
\end{align}
where the matrix $\Sigma_{ij}=\langle s_i s_j\rangle$ is the covariance between observed signals at frequency $i$ and $j$, and a dot means an inner product of vectors. 
In harmonic space, $\Sigma$ is reduced to the observed cross power spectra at multipole $\ell$.
Using this formalism we find the power spectrum of the isolated tkSZ signal, $\widehat C^{\alpha\alpha}_{\ell}$, to be
\begin{align} \label{eq:cILCpow}
\widehat C^{\alpha\alpha}_{\ell} =\frac{\vec{G}\cdot  \Sigma_{\ell}^{-1} \vec{G}}{(\vec{A}\cdot  \Sigma_\ell^{-1} \vec{A}) ( \vec{G} \cdot \Sigma_\ell^{-1}\vec{G})  -(\vec{A} \cdot \Sigma_\ell^{-1} \vec{G}) ^2},
\end{align}
which includes the signal, experimental noise and residual foregrounds.
The resulting spectra, dominated by noise and residual foregrounds, are shown in Fig.~\ref{fig:powerspectra} for our two reference experiments. We note that the constraint of demanding zero response to the kSZ comes at the cost of increasing the variance of the tkSZ map on small angular scales by a factor of $4$ (for the ground based experiment); however, to be conservative we include this constraint in our forecasts below. 
\begin{table}
\centering
\caption{Spectral and spatial dependence of the foregrounds we include in isolating the tkSZ effect. Here, $B_\nu(T)$ is the  spectral radiance of a blackbody at the CIB or dust temperature and we use Refs.~\cite{Dunkley2013,PlanckDust} for the parameters in the table.}
\begin{tabular}[t]{|l|c|c|}
\hline
Component & Spectral dependence & Spatial dependence \\
\hline
CMB +kSZ &  $\propto\nu^3 \mathcal G$ & $C^{\Theta \Theta}_{\ell}$\\
tSZ & $\propto \nu^3 \mathcal Y$ &  $C^{yy}_{\ell}$ \\
CIB &  $\propto\nu^{\beta_\mathrm{CIB}} B_\nu(T_{\rm CIB}) $ & $A_p^\mathrm{CIB} + A_c (\ell / \ell_c)^{\alpha_\mathrm{CIB}} $\\
Radio & $\propto\nu^{\beta_\mathrm{Radio}} $ & $ A_{\rm{Radio}}(\ell/\ell_c)^{\alpha_\mathrm{Radio}}$ \\
Galactic Dust  & $\propto\nu^{\beta_\mathrm{Dust}} B_\nu(T_{\rm Dust})$ &$ A_{\rm{Dust}} (\ell/\ell_c)^{\alpha_\mathrm{Dust}}$ \\
\hline
\end{tabular}
\label{tab:foregrounds}
\end{table}%

\begin{table*}
\centering
\caption{Space-based experimental parameters (based on PICO)}
\begin{tabular}[t]{|l|ccccccccccccccccccccc|}
\hline
Frequency (GHz)   & 21  & 25  & 30  & 36  & 43  & 51  & 62  & 74  & 89  & 107  & 129  & 154  & 185  & 222  & 267  & 321  & 385  & 462  & 554  & 665  & 798  \\
 Beamsize (arcmin)  & 38.4  & 32.0  & 28.3  & 23.6  & 22.2  & 18.4  & 12.8  & 10.7  & 9.5  & 7.9  & 7.4  & 6.2  & 4.3  & 3.6  & 3.2  & 2.6  & 2.5  & 2.1  & 1.5  & 1.3  & 1.1  \\
 Ref. noise ($\mu$Karcmin) & 16.9  & 11.9  & 8.0  & 5.6  & 5.6  & 4.0  & 3.8  & 3.0  & 2.0  & 1.4  & 1.5  & 1.3  & 2.8  & 3.2  & 2.2  & 3.0  & 3.2  & 6.4  & 32.4  & 125.3  & 740.3  \\
\hline
\end{tabular}
\label{tab:pico}
\end{table*}%
\begin{table}
\caption{Ground-based experimental parameters (based on CMB-HD)}
\begin{tabular}[t]{|l|ccccccc|}
\hline
Frequency (GHz)   & 30  & 40  & 90  & 150  & 220  & 280  & 350  \\
 Beamsize (arcmin)  & 1.25  & 0.94  & 0.42  & 0.25  & 0.17  & 0.13  & 0.11  \\
 Ref. noise ($\mu$Karcmin) & 6.5  & 3.4  & 0.7  & 0.8  & 2.0  & 2.7  & 100.0  \\
\hline
\end{tabular}
\label{tab:cmbhd}
\end{table}

\textit{The tkSZ power spectrum---.}
Here we straightforwardly extend the formalism in Ref.~\cite{Ma2002} to compute the tkSZ power spectrum. This is done by replacing the electron density in the kSZ power spectrum with the electron pressure. Thus the tkSZ power spectrum is
\begin{align}
 C^{\alpha\alpha}_{\ell}  =   & \frac{1}{2}\int \frac{\mathrm{d \chi}}{\chi^2} (RaH)^2f^2 \int\frac{\mathrm{d}^3k'}{(2\pi)^3}  \biggl(P_{pp}(|\mathbf{k}-\mathbf{k'}|,z) \nonumber  \\ &    \times P^{\mathrm{lin}}_{\delta\delta}(k',z)\frac{k(k-2k'\mu')(1-{\mu'}^2)}{{k'}^2(k^2+{k'}^2-2kk'\mu')} \biggr) \bigg\vert_{k=\frac{\ell}{\chi}},
\end{align}
where $H$ is the Hubble parameter, $f$ is the logarithmic derivative of the growth factor ($ \mathrm{d}\log D / \mathrm{d}\log a$), $\mu' = \hat{\mathbf{k}}\cdot \hat{\mathbf{k}}'$ and $P^{\rm lin}_{\delta \delta}$ is linear matter power spectrum. We use the halo model \cite{Cooray2002} to compute the non-linear pressure power spectrum $P_{pp}$, using a modified version hmvec code \footnote{hmvec was made by Mathew Madhavacheril and is available at \url{https://github.com/msyriac/hmvec} } and with the cluster pressure profiles from Ref.~\cite{Battaglia2012}. In addition we compute the cross power spectrum with the kSZ effect, using the electron density profiles from Ref.~\cite{Battaglia2016}~(due to the velocity dependence there is no cross correlation with tracers like the tSZ or CIB).

One may wonder if we are able to directly observe tkSZ power spectrum or tkSZ-kSZ cross power spectrum due to the specific frequency dependence of the tkSZ effect.
However, we find that measuring these correlations is difficult as the auto-spectrum is tiny and, at large scales, the cross spectrum is limited by the CMB noise from the primary CMB (which cannot be separated from the kSZ). The auto and cross power spectra are shown in Fig.~\ref{fig:powerspectra}. We see that the signal is small and beyond the detection capabilities of upcoming experiments.  

\textit{The galaxy-galaxy-tkSZ bispectrum---.}
Several methods have been studied, and recently used, for detecting the kSZ effect from a sample of galaxies. These include stacking differences of the CMB temperature at locations of objects~\cite{Hand:2012ui,DeBernardis:2016pdv,Li2018,Ade:2015lza}, as well as performing a velocity-weighted stack of CMB temperatures on the objects, in which the velocities are obtained from a three-dimensional map of structure \cite{Schaan2016,Ma:2017ybt}. It was recently shown by Ref.~\cite{Smith:2018bpn} that these methods are all effectively measures of the galaxy-galaxy-kSZ bispectrum, which is a sensitive probe of a range of cosmological parameters~\cite{Munchmeyer2018,Pan2019}. The velocity dependence of the tkSZ effect means that we can utilise the same machinery to study the galaxy-galaxy-tkSZ bispectrum, which is the main result of this paper.
Following Ref.~\cite{Smith:2018bpn}, we use a simplified `snapshot' geometry to approximate cross correlations with galaxy surveys. We use this geometry to simplify the light cone evolution effects. Here we model the universe as a periodic 3D box with comoving side length $L$ at a snapshotted redshift $z_*$ and use $*$ to denote quantities at this snapshotted time.
We also work in the flat-sky approximation, assuming the sky to be periodic with angular side length $L/\chi_*$.

We decompose the observed galaxy density field, $g(\mathbf x)$ and tkSZ map, $\alpha({\bm n})$, into Fourier components as $g(\mathbf k) \equiv \int \mathrm{d}^3\mathbf{x} g(\mathbf{x})e^{-i\mathbf{k}\cdot\mathbf{x}}$, and $\alpha({\bm \ell}) \equiv \int \mathrm{d}^2{\bm n} \alpha({\bm n} )e^{-i{\bm n} \cdot {\bm \ell}}$.
Symmetries simplify the momentum dependence of the bispectrum and we find~\cite{Smith:2018bpn}
\begin{align}
	\langle g(\mathbf k)g(\mathbf k')\alpha({\bm \ell})\rangle'=iB_{gg\alpha}(k,k',\ell,k_r),
\end{align}
where the prime on the bracket implies that we omitted $(2\pi)^3 \delta^{(3)}\left(\mathbf k+\mathbf k'+ {{\bm \ell}/{\chi_*}}\right)$.
Note that $-k'_r=k_r$ is the projection of the Fourier momenta onto the $z$ axis. We measure the detectability via the Fisher information of $B_{gg\alpha}$, which is given by
\begin{align}\label{Def:fisher}
	&F =\frac{V}{2}\int  \frac{ \mathrm{d}k_L \mathrm{d}k_S \mathrm{d}k_{Lr}}{8\pi^3}k_Lk_S \chi^2_{*} \nonumber \\ &  \times \left(\frac{B_{gg\alpha}(k_L,k_S,\ell,k_{Lr})B_{gg\alpha}^*(k_L,k_S,\ell,k_{Lr})}{P_{gg}^{\rm tot}(k_L)P_{gg}^{\rm tot}(k_S) \widehat C^{{\alpha} {\alpha}}_{\ell}}\right)_{\ell=k_s\chi_*}
	\end{align}
where $V\equiv L^3$ and $\widehat C^{{\alpha} {\alpha}}_{\ell}$ is the measured angular power spectrum of the tkSZ effect from the cILC, Eq.~\eqref{eq:cILCpow}, and $P^{\rm tot}_{gg}(k)$ is the observed galaxy power spectrum including the shot noise. Ref.~\cite{Smith:2018bpn} showed that the Fisher information is dominated by squeezed bispectrum configurations, i.e. $k_{L}\ll k_{S}\sim \ell/\chi_*$. With the approximations of Ref.~\cite{Smith:2018bpn}, that the squeezed bispectrum can be approximated by the tree level bispectrum with the linear power spectra replaced by the non linear power spectra, we find
\begin{align}
B_{gg\alpha} \approx  \frac{   R_*  k_r }{ \chi_*^2}
	  	\left[\frac{P_{gv}(k')}{k'}P_{gp}(k)-\frac{P_{gv}(k)}{k} P_{gp}(k')\right],\label{tkSZ:bis}
\end{align}
where $P_{XY}\equiv \langle X(\mathbf k)Y(\mathbf k') \rangle'$ are the cross power spectra of $X$ and $Y$. To compute the non-linear $P_{gp}(k)$ we use the halo model \cite{Cooray2002}, the galaxy HOD \cite{Zheng2005,Berlind2002} from Refs.~\cite{Leauthaud2011,Leauthaud2012} and the cluster pressure profiles from Ref.~\cite{Battaglia2012}. With this approximation, Eq.~\eqref{Def:fisher} is then reduced to
\begin{align}
		F &=
		\frac{V   R_*^2  }{12\pi^3\chi_*^2}
	  	\int dq_L \frac{ q_L^2P^2_{gv}(q_L) }{P^{\rm tot}_{gg}(q_L)}
	  	\int dq_S
	  	\frac{q_SP^2_{gp}(q_S) }{P^{\rm tot}_{gg}(q_S) \widehat C^{{\alpha}{\alpha}}_{k_S\chi_*}}\label{fisher:sqlimit}.
	  \end{align}
The main goal of this paper is to discuss the measurability and utility of the tkSZ effect, and we evaluate Eq.~\eqref{fisher:sqlimit} to accomplish this. 
We consider how our two reference experiments can be combined with an upcoming DESI-like spectroscopic survey~\cite{Aghamousa:2016zmz} to measure the tkSZ bispectrum. We assume the mean redshift of the galaxy survey
is 0.75, that
the volume of overlap
with the CMB survey is 116 Gpc$^3$, the galaxy bias is
1.51 and the observed galaxy number
density $1.7\times 10^{-4}$ Mpc$^{-3}$.
Combining this setup with the isolated
tkSZ power spectrum~\eqref{eq:cILCpow} we evaluate $\sqrt{F}$ of Eq.~\eqref{fisher:sqlimit} to estimate the detectability of the tkSZ bispectrum. We note that due to their differing velocity dependence, the CIB and tSZ will not bias the measurement, as with the kSZ bispectrum.

\begin{figure}[t]
  \centering
  \includegraphics[width=0.48\textwidth]{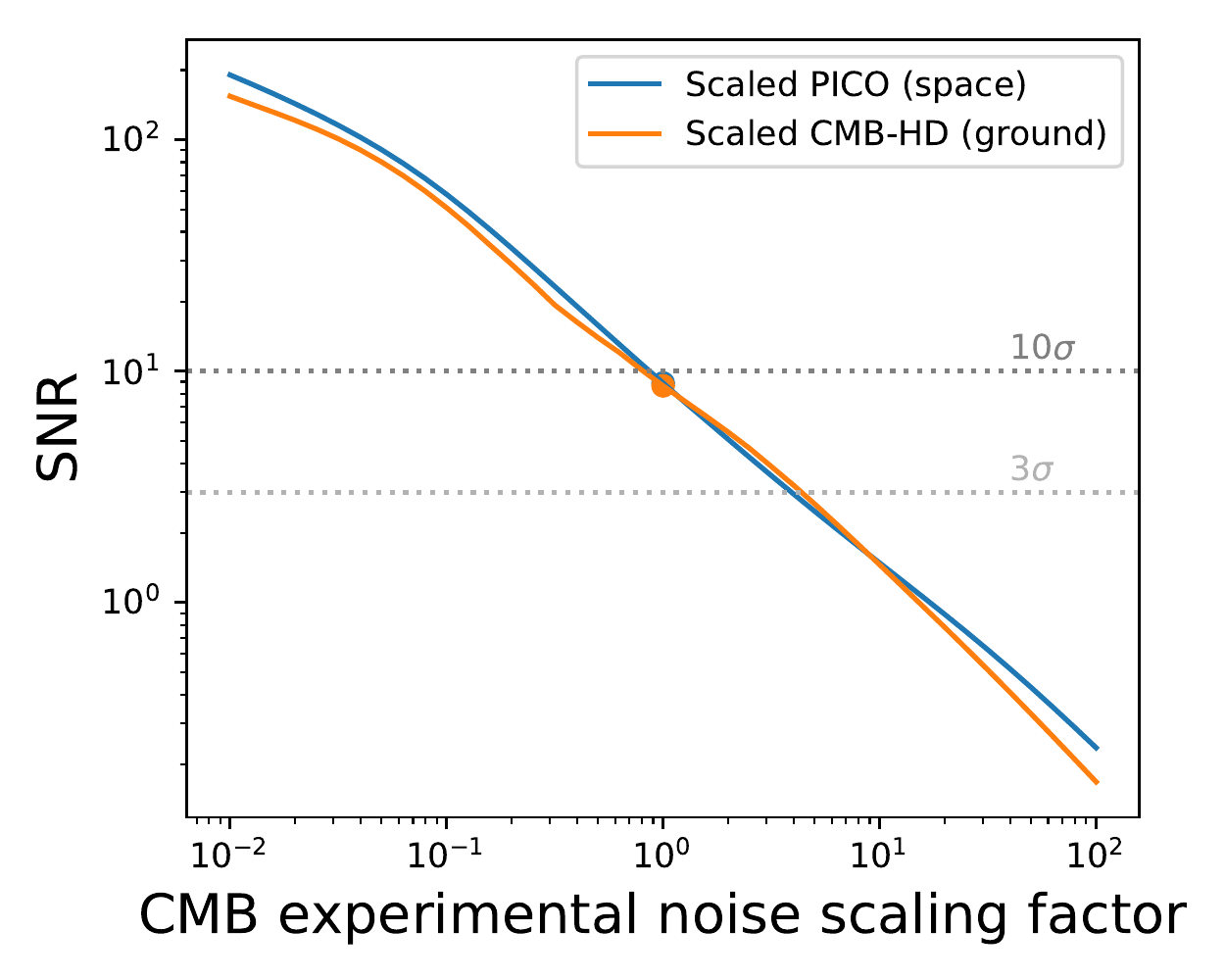}
\caption{Signal to noise ratio for the galaxy-galaxy-tkSZ as would be obtained using methods similar to current methods to constrain the kSZ effect.  Using Eq.~\eqref{fisher:sqlimit} we find that the sensitivities of the nominal ground-based and space-based experiments could yield detection significances of 8$\sigma$. On the horizontal axis we scale these nominal map level instrumental sensitivities at all frequencies by the amount shown, in order to demonstrate the impact on the SNR.
}
\label{fig:SNR}
\end{figure}

We find that each of the two experiments would be able to detect the galaxy-galaxy-tkSZ bispectrum at the $ \sim 8 \sigma$ level. We note that the two experiments achieve this SNR from slightly different scales, with the CMB-HD experiment gaining information from much smaller scales. In Fig.~\ref{fig:SNR} we explore the detectability as the experimental noise level in all channels is scaled by the same factor, finding that a $3\sigma$ detection would still be possible from a ground- or space-based survey if the map-level noise were increased by a factor of $\sim 4$ compared to CMB-HD and PICO respectively. In general, a key requirement to detect this signal is a wide range of frequencies with low noise. This is essential to separate out the foregrounds and deproject the kSZ signal. 
For an upcoming ground-based survey with more modest angular resolution than CMB-HD, namely CMB-S4~\footnote{CMB-S4 parameters from \url{https://cmb-s4.org/wiki/index.php/Survey_Performance_Expectations}} 
used together with CCAT-prime for the higher frequencies \cite{2019arXiv190810451C}, we find a marginal detection prospect of $2.1\sigma$.

In the more distant future, this effect could be detected at a very high significance. We find that the cosmic variance limit SNR (with an $\ell_{\rm max} = 5000$) is 1340, assuming the covariance matrix only has Gaussian contributions and only including squeezed limit contributions. Whilst current experiments will be able to detect the kSZ effect at higher significance, the cosmic variance limit for the tkSZ is larger; for example with an $\ell_{\rm max} = 5000$ the tkSZ can be studied at twice the precision of the kSZ effect.

\textit{Conclusion---.}
In this work we highlighted the use of the tkSZ effect as a new cosmological observable. This signal can be isolated from other components due to its unique spectral signature and would be detectable by the envisioned PICO and CMB-HD experiments.
Whilst we have focused on outlining the origin and detectability of this effect, there are several interesting potential applications. Perhaps most interestingly, we can use the galaxy-galaxy-tkSZ bispectrum, whose detectability we discussed above, to reconstruct the large scale velocity field. This can be converted to a measurement of the large scale density field, via the continuity equation, with noise that scales as $k$ and is ideal to probe scale dependent bias for constraining primordial non-Gaussianity~\cite{Dalal:2007cu}. This provides significant advantages to the shot noise on large scales obtained directly from galaxy survey measurements of the large scale density fields. 

Using tkSZ tomography offers a few benefits, in the long term, compared to standard kSZ tomography. Firstly, the tkSZ approach does not suffer from the optical depth degeneracy. This problem arises for kSZ tomography as the distribution of electrons is uncertain and means that velocity reconstruction methods have an unknown normalization. This is not the case for the tkSZ, since it instead depends on the cluster pressure profile, which can be determined by tSZ measurements to a much higher precision than the electron distribution. Secondly, the tkSZ spectrum is dominated by the signal of interest. For kSZ measurements the large scales are dominated by the CMB and at small scales there is a significant contribution to the power spectrum from reionization. In the future this will act as a source of noise for kSZ tomography that cannot be removed. The tkSZ effect is immune from this. These features can be seen from the low-noise behavior of the forecasted signal to noise ratio in Fig.~\ref{fig:SNR} in which the detectability of our signal continues to increase rapidly as the instrumental noise is reduced. In the cosmic variance limit this effect can be detected at more than 1000 $\sigma$.

Finally we note that this work has focused on the galaxy-galaxy-tkSZ bispectrum. However in the longer term, the tkSZ signal can be isolated and be measured at the power spectrum through cross correlations with the kSZ and, more excitingly but more distant, through its auto-spectrum. The tkSZ auto-spectrum can be directly measured on large scales, which is impossible for the standard kSZ effect due to the presence of the primary CMB, and thus can directly probe the pressure velocity power spectra. We defer a thorough discussion of this to the future. Nevertheless, in this paper we have demonstrated a new cosmological signal that is within the reach of currently envisioned CMB experiments. 

\begin{acknowledgments}
We would like to thank Anthony Challinor, Jens Chluba, Colin Hill, Neelima Sehgal, and Kendrick Smith for helpful discussions.
AO is suppoted by Japan Society of Promotion of Science Overseas Research Fellowships.
AvE would like to thank the Kavli Institute for Cosmology Cambridge for their support and hospitality for his visit under their Visiting Scholars program, during which this work was initiated. The ``hmvec" halo model code made by Mathew Madhavacheril was used in this analysis. WRC acknowledges support from the UK Science and Technology Facilities Council (grant number ST/N000927/1). 
\end{acknowledgments}

\bibliography{reference.bib}{}
\bibliographystyle{unsrturl}

\end{document}